\documentclass[preprint,pra,showpacs,floatfix]{revtex4}
\usepackage{amsmath}
\usepackage{amsfonts}
\usepackage{amssymb}
\usepackage{graphicx}

\begin{document}

\title{A Model for Non-Cancellation of Quantum Electric Field Fluctuations}

\author{Victor Parkinson }
\email{victor@cosmos.phy.tufts.edu}
\author{L. H. Ford }
\email{ford@cosmos.phy.tufts.edu}
\affiliation{Institute of Cosmology, Department of Physics and Astronomy\\
    Tufts University, Medford, MA 02155, USA}
    
\begin{abstract}
A localized charged particle oscillating near a reflecting boundary is considered as a model for
non-cancellation of vacuum fluctuations. Although the mean velocity of the particle is
sinusoidal, the velocity variance produced by vacuum fluctuations
 can either grow or decrease linearly in time, 
depending upon the product of the oscillation frequency and the distance to the boundary.
This amounts to heating or cooling, arising from non-cancellation of electric field fluctuations,
which are otherwise anticorrelated in time.
 Similar non-cancellations arise in quantum field effects in
 time-dependent curved spacetimes.
We give some estimates of the magnitude of the effect, and
discuss its potential observability.  We also compare the effects of vacuum fluctuations 
with the shot noise due to emission of a finite number of photons. We find that the two effects
can be comparable in magnitude, but have distinct characteristics, and hence could be 
distinguished in an experiment.
\end{abstract}
    
\maketitle

\baselineskip=13pt

\section{Introduction}
\label{sec:intro}

Consider a localized charged particle coupled to quantum electromagnetic
field fluctuations in the vacuum state.  We will treat it as a classical particle,
but more generally it can be viewed as
a quantum particle in a wavepacket state sharply peaked in
space. Because the vacuum is the state of lowest energy of the quantum
field, the particle cannot, on average, acquire energy from the electromagnetic
field. This does not prevent energy fluctuations which are within the
limits set by the energy-time uncertainty principle. The particle can
acquire additional energy from an electric field fluctuation, but the
energy must be surrendered on a timescale inversely proportional to the
magnitude of the energy. Energy conservation is enforced by temporally
anticorrelated electric
field fluctuations, which are guaranteed to take back the energy within
the allowed time. Thus on the average, neither the particle nor the quantum field
gains energy.

This holds in any static situation, including one where reflecting
boundaries are present. Although classical image charge effects can be
present, no net energy may be extracted from the vacuum. A model with
a charge maintained at fixed mean distance from a plane mirror was
treated in Ref.~\cite{YF04}. Switching on the effect of the mirror can cause
the particle's mean squared velocity to either increase or decrease,
but after transients have died away, it approaches a constant. This 
need not be the case in a time-dependent situation, which will be the
topic of this paper. The cause of the time-dependence may be a source
of energy, so it is now possible for the particle's energy to either grow or
decrease in time. However, one may also view the time-dependence as
upsetting the anticorrelated fluctuations which are present in a static 
situation.  In the static case, the anticorrelated fluctuation takes exactly
the amount of energy obtained by the particle in a previous fluctuation.
The time-dependence may either enhance or suppress the magnitude of the
the second fluctuation, resulting in either a decrease or increase, respectively,
of the particle's energy. We will see both possibilities illustrated in the model
discussed in Sect.~\ref{sec:model}. 

Examples of non-cancellation of field fluctuations arise in cosmology. One
is Brownian motion of charged particles in an expanding 
universe~\cite{Bessa09}. Other examples were discussed in 
Refs.~\cite{WKF07,FMNWW10,WHFN11}, where it was argued that quantum
stress tensor fluctuations during inflation can lead to density and gravity wave
perturbations which depend upon the total expansion during inflation. In
the present paper, we consider a simple flat space model which is of interest
both in its own right, and as an analog model for effects in curved spacetime.
Lorentz-Heaviside units with $c=\hbar=1$ will be used. 

\section{The Model}
\label{sec:model}

\subsection{Formulation and Calculations}

Our model consists of a particle of mass $m$ and electric charge $q$ undergoing
bounded, non-relativistic motion in a direction normal to a perfectly reflecting plane mirror.
We take this to be the $z$-direction, and write
 \begin{equation}
z(t) = d + A\, f(t) \,,                      \label{eq:traj}
\end{equation} 
 where $d$ is the mean distance to the mirror, $A>0$ is  the amplitude
 of the motion, and $f(t)$ is a dimensionless function which we later take to be
 sinusoidal. We require $z(t) > 0$ for all $t$ and $|\dot{z}(t)| = A\, |\dot{f}(t)| \ll 1$ 
  We assume that the components of the particle's velocity satisfy a Langevin
  equation, 
  \begin{equation}
  \dot{v}_i = \frac{q}{m}\, E_i (\mathbf{x},t) \,,                      \label{eq:eom}
\end{equation} 
where $\mathbf{x} = \mathbf{x}(t)$ is the spatial location of the particle at time $t$.
Here ${\bf E}$ is the total electric field, including both a classical applied field,
including possible image charge effects,
and the quantized electric field. This is the usual equation of motion for a 
non-relativistic charged particle when magnetic forces are neglected. Our
key assumption is that it may be used in the presence of a fluctuating electric
field. For now we ignore dissipation effects,
which have been discussed in Refs.~\cite{HL06,WHL08}. We will treat dissipation
by emitted radiation in Sect.~\ref{sec:rad}. Note that an alternative to moving the
charge with the mirror fixed is to move the mirror, or to use a charge moving at
constant speed near a corrugated mirror. The latter strategy was first used by
Smith and Purcell~\cite{SP} to create radiation, and is the basis of the free electron
laser.  
  
With the initial condition $v_i(t_0) = 0$, we may integrate the Langevin equation
and then take expectation values in the electromagnetic field vacuum state to
write the variance in $v_i$ as a double time integral of the electric field correlation
function:
\begin{equation}
\langle{\Delta v_i^2(t)}\rangle = 
\frac{q^2}{m^2}\; \int_{t_0}^t dt_1  \int_{t_0}^t  dt_2 \,
[\langle{E}_i({\mathbf x}_1,t_1)\;{E}_i({\mathbf x}_2,t_2)\rangle -  
 \langle{E}_i({\mathbf x}_1,t_1)\rangle\; \langle{E}_i({\mathbf x}_2,t_2)\rangle]\,.
\end{equation} 
Here ${\mathbf x}_1 = \mathbf{x}(t_1)$ and  ${\mathbf x}_2 = \mathbf{x}(t_2)$, the
spatial locations of the particle at times $t_1$ and $t_2$, respectively.
Any classical part to the electric field will cancel in the correlation function.  For now, we
focus on the quantum part of the electric field, for which $\langle{E}_i({\mathbf x},t)\rangle = 0$. 
We are interested only in the effect of the boundary, as the empty space correlation
function will not produce any growing terms in $\langle{\Delta v_i^2(t)} \rangle$. The
quantum electric field correlation function may be written as a sum of an empty space part
and a boundary correction. We drop the former and write
\begin{equation}
\langle{\Delta v_i^2}\rangle = \frac{q^2}{m^2} \int_{t_0}^t   dt_1  \int_{t_0}^t  dt_2 \langle 
E_i(\mathbf{x}_1,t_1)E_i(\mathbf{x}_2,t_2)\rangle_b \,,
\end{equation}
where the subscript b indicates the boundary correction to the two-point function.
These corrections may be found
by the method of images, and are~\cite{BM69} 
\begin{equation}
\langle E_x(\mathbf{x}_1,t_1)E_x(\mathbf{x}_2,t_2)\rangle_b = 
- \frac{\tau^2 + (z_1+z_2)^2}{\pi^2 [\tau^2 - (z_1+z_2)^2 ]^3}
\label{eq:Ex}
\end{equation}  
 for a transverse direction, and
\begin{equation} 
  \langle E_z(\mathbf{x}_1,t_1)E_z(\mathbf{x}_2,t_2)\rangle_b = 
  \frac{1}{\pi^2 [\tau^2 - (z_1+z_2)^2 ]^2}           \label{eq:Ez}
\end{equation}
for the longitudinal direction, where $\tau = t_1 - t_2$, and $z_1 = z(t_1)$, ect. 
Here we assume that the particle does not move far compared to the distance to the
mirror, and have equated the coordinates in the transverse directions, $x_1=x_2$
and $y_1=y_2$.
Note that, for example, $\langle v_x\, v_z \rangle_b =0$,
so there will be no correlation between the random motion in the transverse and
longitudinal directions.  

Next we assume that $|A\,f(t)| \ll d$, and Taylor expand the two-point functions to second
order in $A$. The integrand for the longitudinal variance becomes
\begin{eqnarray}
& &\frac{1}{(\tau^2 - \{ 2d + A[f(t_1) + f(t_2)] \}^2)^2} \approx \nonumber  \\
& & \frac{1}{(\tau^2 - 4d^2)^2} + \frac{8d}{(\tau^2 - 4d^2)^3}A[f(t_1)+f(t_2)]  + \frac{2(\tau^2 + 20d^2)}{(\tau^2 - 4d^2)^4}A^2[f(t_1)+f(t_2)]^2 \,. \label{eq:expand}
\end{eqnarray}
We are seeking  contributions to $\langle{\Delta v_i^2}\rangle$ which grow in time. 
 The zeroth order term describes the case of a stationary charge, which was treated
 in Ref.~\cite{YF04}, and gives a constant contribution. The first order term yields a
purely oscillatory function when $f(t)$ is sinusoidal.  Thus we omit both of these terms 
and focus on the second order term. Note that the $\tau$-dependent part of this
term may be written as a total derivative
\begin{equation}
F_z(\tau) \equiv \frac{2(\tau^2 + 20 d^2)}{(\tau^2 - 4d^2)^4} \equiv  \frac{d^4}{d\tau^4} G(\tau)
= \frac{\partial^2}{\partial t_1^2} \frac{\partial^2}{\partial t_2^2} G(\tau) \,.
\end{equation}
The function $G(\tau)$ may be expressed in terms of logarithmic functions, but we
will not need its explicit form, beyond the fact that it has only a logarithmic singularity
at $\tau = 0$. 

Now we assume that $f(t)$ and its first three derivatives vanish in the past and future.
This allows us to integrate over all $t_1$ and $t_2$, and to perform integrations by
parts with no boundary terms. Thus we may write
\begin{equation}
\int_{-\infty}^\infty   dt_1    dt_2\,F_z(\tau)\, [f^2(t_1) +f^2(t_2)]
  = \int_{-\infty}^\infty   dt_1   dt_2 \,G(\tau)\,
  \frac{\partial^2}{\partial t_1^2} \frac{\partial^2}{\partial t_2^2}   [f^2(t_1) +f^2(t_2)]  = 0\,.
\end{equation}
This implies that only the cross term in the last term in Eq.~(\ref{eq:expand}) can give
a nonzero contribution. Now we may write 
\begin{equation}
\langle{\Delta v_z^2}\rangle = 
\frac{2}{\pi^2}\frac{q^2}{m^2}A^2 \int_{-\infty}^\infty\;\int_{-\infty}^\infty\; dt_1\;dt_2\;F_z(\tau)\,f(t_1)  
\, f(t_2) \,.             \label{eq:vz}
\end{equation}

Next we adopt a specific form for $f(t_1)$, which is $f(t_1) = \sin(\omega t_1)$
for $0 \alt t_1 \alt t$ and $f(t_1) =0$ for $t_1 \alt 0$ and $t_1 \agt t$. The approximate
signs indicate that $f$ should fall smoothly to zero at the end points of the interval.
This describes a charge which oscillates sinusoidally at angular frequency $\omega$
for a time $t$. This sinusoidal motion could be driven by a classical electric field
of the form $E_z^{\rm cl}(t) = -E_0\, \sin(\omega t)$, in which case
 \begin{equation}
A = \frac{q\, E_0}{m\, \omega^2} \,.       \label{eq:A}
\end{equation}
The integration in Eq.~(\ref{eq:vz}) is effectively over a square of side $t$.
Next, we change integration variables to $\tau$ and $u=t_1 + t_2$. Because $F_z(\tau)$
falls to zero rapidly if $|\tau| \gg d$, and because we assume $t \gg d$, 
the integration on $\tau$ may be taken over an infinite range. However, the $u$ 
integration is restricted to a finite interval:
 \begin{equation}
\langle{\Delta v_z^2}\rangle = 
\frac{1}{2\pi^2}\frac{q^2}{m^2}A^2 \int_{0}^{2t} du  \int_{-\infty}^\infty\; d\tau 
F_z(\tau)[\cos(\omega\tau) - \cos(\omega u)] \,.
\end{equation}
The integral of the $\cos(\omega u)$ term will generate an entirely oscillatory contribution,
which may be ignored compared to the linearly growing term, so we may write
\begin{equation}
\langle{\Delta v_z^2}\rangle  \approx
 \frac{2}{ \pi^2}\, \frac{q^2}{m^2} \,A^2 \; t  \left[ \int_{-\infty}^\infty\; 
\frac{(\tau^2 + 20d^2)}{(\tau^2 - 4d^2)^4}\, \cos(\omega \tau) \;d\tau \right]   \,.
\label{eq:zInt}
\end{equation}

At this point, it is useful to note that $\tau$ should have a small, negative imaginary part
in Eqs.~(\ref{eq:Ex}) and (\ref{eq:Ez}). This arises because these two-point functions
are expressible as integrals of the form
\begin{equation}
\int_0^\infty d\omega \, \omega^3 \, {\rm e}^{-i \omega \tau} \,,
\end{equation}
which are absolutely convergent if ${\rm Im}(\tau) <  0$. We can implement this condition
by replacing $\tau$ by $\tau - i \epsilon$ in Eq.~(\ref{eq:zInt}), where $\epsilon$ is a small
positive real number. We can write the denominator in the integrand as 
\begin{equation}
[(\tau -i \epsilon)^2 - 4d^2]^4 = (\tau  -i \epsilon+ 2d)^4(\tau  -i \epsilon- 2d)^4 \,,
\end{equation}
revealing  that there are two  fourth-order poles in the upper half-plane  at 
$\tau = \pm 2d  + i \epsilon$.  Next we write $\cos(\omega \tau)$ in terms of complex
exponentials. The $\tau$ integration is along the real axis, so the $ {\rm e}^{-i \omega \tau}$ 
term gives no contribution when the contour is closed in the lower half-plane. The
$ {\rm e}^{i \omega \tau}$ term yields the residues of the two poles when the contour is
closed in the upper half-plane. The sum of the residues is a real function.

\subsection{Key Results}

The result of the evaluations of the longitudinal velocity variance,
 after using Eq.~(\ref{eq:A}),  is 
\begin{equation}
\langle{\Delta v_z^2}\rangle =   \frac{q^4\, E_0^2}{16 \pi m^4 \,d} \,R_z \, t \,,
\label{eq:vz2}
\end{equation}
where
\begin{equation}
R_z = \frac{1}{2 \xi^4} [(3 -5 \xi^2)\, \sin(2 \xi) + 2\xi \,(\xi^2-3) \, \cos(2 \xi)] \,,
\end{equation}
and $\xi = \omega \, d$.

\begin{figure}
  \begin{center}
    \includegraphics[scale=0.5]{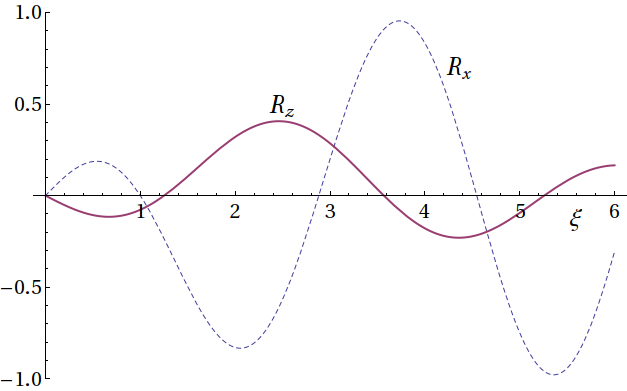}
  \end{center}
  \caption{ This graph shows the dimensionless quantities $R_z$ and $R_x$, 
  which are proportional to the rates of
  change of velocity variance in the longitudinal direction and a transverse direction,
  respectively. Here $\xi = \omega\, d$.}
  \label{fig:R-graph}
\end{figure}

 The same mathematical technique holds for the transverse direction; only the precise 
 form of the integrand changes. Let $F_z \rightarrow F_x$, where
 \begin{equation}
F_x(\tau) = -\frac{4(40d^4 + 34d^2\tau^2 + \tau^4)}{[(\tau -i \epsilon)^2 - 4d^2]^5}\,.
\end{equation}
In this case, there are two fifth-order poles in the upper half-plane, but otherwise
the evaluation procedure is the same.
Now the velocity variance in the $x$-direction, which is also the mean
squared velocity in this direction, is found to be
\begin{equation}
\langle{\Delta v_x^2}\rangle = \langle{ v_x^2}\rangle =
 \frac{q^4\, E_0^2}{16 \pi m^4 \,d} \, R_x \, t        \,,\label{eq:vx2}
\end{equation}
where
\begin{equation}
R_x = \frac{\xi^2 -1}{4 \xi^4} [(4 \xi^2 -3)\, \sin(2 \xi) + 6\xi \, \cos(2 \xi)] \,,
\end{equation}
Note that the $R_i$, which are dimensionless, are proportional to the
rate of change of the corresponding velocity variance:
\begin{equation}
R_i(\xi)= \frac{16 \pi m^4 d}{ q^4 E_0^2} \;\frac{d\langle \Delta v_i^2 \rangle}{dt}\,.
\end{equation}
These quantities are  illustrated in Fig.~\ref{fig:R-graph}.

 Of significant interest here is that for both the longitudinal and transverse components, 
 the coefficient of the time dependence of $\langle \Delta v_i^2 \rangle$ can be either 
 positive or negative, depending on the frequency of the oscillation and distance 
 to the mirror. These results can be interpreted in terms of non-cancellation of previously
 anticorrelated electric field fluctuations. When there is linear growth, the fluctuations
 are adding energy to the particle on average. Similarly, a linear decrease signifies
 that they are removing energy, which could be described as a  ``cooling mode" .
 The latter effect can only go so far, and  at some point our approximation of localized particles
 would break down.
 
It is also of interest to examine the low and high frequency limits of the above results. 
At low frequency, $\xi \ll 1$, we have 
\begin{equation}
\langle{\Delta v_x^2}\rangle \sim -2\langle{\Delta v_z^2}\rangle \sim
\frac{q^4\, E_0^2\, \xi}{30 \pi m^4 \,d} \,t \,, 
\end{equation}  
 and at high frequency,    $\xi \gg 1$, 
 \begin{equation}
\langle{\Delta v_x^2}\rangle \sim \frac{q^4\, E_0^2\, t}{16 \pi m^4 \,d} \, \sin(2 \xi)\,,
\quad
\langle{\Delta v_z^2}\rangle \sim \frac{q^4\, E_0^2\, t}{16 \pi m^4 \,d\,\xi} \, \cos(2 \xi)\,.
 \end{equation} 
 Note that the effect tends to be larger in a transverse direction than in the longitudinal
 direction, especially at high frequencies. 
  
 Next we wish to make some estimates of the magnitude of the heating or cooling effect.
 We do this by defining a change in effective temperature for the $i$-direction, $\Delta T_i$,
 by 
 \begin{equation}
\frac{1}{2}m \langle \Delta v_i^2 \rangle = \frac{1}{2}k_B \Delta T_{i}\,,
\end{equation}
where $k_B$ is Boltzmann's constant.
Strictly speaking, this is not a real temperature, since it is not isotropic, but it is a useful
measure of the size of the effect. From either of Eqs.~(\ref{eq:vz2}) or (\ref{eq:vx2}), we find  
 \begin{equation}
\Delta T_{i} =  \frac{q^4\, E_0^2}{16 \pi k_B m^3 \,d} \, R_i \, t  \,.
\end{equation}  
This may be expressed as
 \begin{equation}
\Delta T_{i}  \approx   10^{-8}\,K\; \left(\frac{I}{1\,W/cm^2}\right)\left(\frac{1 \mu m}{d}\right)\left(\frac{t}{1\,s}\right)\; R_i\, ,
\end{equation}
 where we have replaced $E_0^2/2$ by $I$, the power per unit area in a plane electromagnetic
 wave with peak electric field $E_0$.  We have also set $q=e$, the electronic charge.
 
 Our approximation of a perfectly reflecting plate
 should hold both for modes whose wavelength is of order $d$ and at angular frequencies
 of order $\omega$. Note that $\xi = 2 \pi d/\lambda$, where $\lambda$ is the wavelength
 of the driving field. From Fig.~\ref{fig:R-graph}, we see that $R_z$ reaches its maximum
 value of about $0.5$ at $\xi \approx 2.5$ and $R_x$ first reaches its maximum of about
 $1.0$ at $\xi \approx 4$. Both of these correspond to $\lambda > d$. If $d \agt \lambda_P$,
 the plasma wavelength of the metal in the plate which can be in the range of $0.1 \mu m$,
 then the perfect reflectivity assumption should be valid.  Ultimately, whether this effect
 can be measured in  a realistic experiment  depends upon the sensitivity of temperature
 measurements, the power intensity $I$ of the driving field which can be used, and the
 time $t$ which can be achieved. On the latter point, it is possible that planar Penning
 traps will be able to achieve very long coherence times with single 
 electrons~\cite{GG10}.
 
 As noted earlier, $\langle{\Delta v_x^2}\rangle = \langle{v_x^2}\rangle$ because
 the mean transverse velocity vanishes, $\langle{v_x}\rangle = 0$. Thus the increased drift   
 in the transverse directions when $\langle{v_x^2}\rangle >0$ is a signature of this effect.
 When $\langle{v_x^2}\rangle < 0$ due to the shift in electromagnetic vacuum fluctuations,
 then we need to interpret the effect as a reduction in mean squared transverse velocity,
 with a positive contribution coming from other effects, such as
 quantum uncertainty in speed, classical thermal effects, or shot noise (to be discussed in
 Sect.~\ref{sec:shot}). This reduction is closely related to the phenomenon of 
 negative energy density
 in quantum field theory, whereby it is possible to reduce to local energy density below 
 the vacuum level with either boundaries or quantum coherence effects~\cite{F09}.
 
 In the longitudinal direction, there is a nonzero mean velocity given by the response to
 the classical driving force. The time averaged square of this velocity is
 \begin{equation}
\langle v_z^2 \rangle_c = \frac{1}{2}\left( \frac{qE_0}{m\omega}\right)^2 = 
\frac{1}{2}(A\omega)^2 \,.
\end{equation}
 It is of interest to compare this quantity with the quantum variance given by 
 Eq.~(\ref{eq:vz2}), and write
 \begin{equation}
\frac{\langle \Delta v_z^2 \rangle}{\langle v_z^2 \rangle_c}  = 
\frac{q^2 \xi^2 \,R_z\, t}{8\pi m^2 d^3} = 0.16 \xi^2 R_z \, 
\left(\frac{1\mu m}{d}\right)^3 \left(\frac{t}{1\,s}\right)\,.
\end{equation}  
 Given that we expect $\xi \sim R_z \sim O(1)$, it is possible for the random motion
 produced by  electric field fluctuations to approach a significant fraction of
  the classical motion. 
  
  So far, we have treated the charges as classical point particles, but the same
  conclusions about changes in $\langle \Delta v_i^2 \rangle$ should hold for
  quantum particles if they are localized in space on scales small compared to
  $d$, the distance to the mirror. Ehrenfest's theorem tells us that classical equations
  of motion hold, on average, in quantum mechanics. We are concerned here
  with changes in the variance of the particles' momentum, which follow
  from momentum conservation considerations.

\section{Radiation and Shot Noise}  
 \subsection{Radiated Power}  \label{sec:rad}
  
In this subsection, we will address the dissipation effect due to emitted electromagnetic
radiation. First, we determine the average power radiated by our 
oscillating charge. It acts as an electric dipole, and so we may use the far-zone field formulas, with the method of images to obtain the field 
components. (This problem has been extensively studied in the literature. See, for example,
Ref.~\cite{LA10} for further references and a detailed treatment of the near-zone fields.)
What follows are the non-zero field components at a point of observation P 
located in the $x-z$ plane, where $r_1$ is the distance from P to the real dipole, and $r_2$ the 
distance from P to the image dipole. We have assumed that P is far enough away that both 
$r_1$ and $r_2$ have approximately the same polar angle $\theta$. First, the z-component:
\begin{equation}
E_z = \frac{\sin^2 \theta}{4\pi}\, p_e\, \omega^2\left(
\frac{e^{i\omega r_1}}{r_1} +
 \frac{e^{\omega r_2}}{r_2} \right)
\end{equation}
where $p_e$ is the peak value of the oscillating electric dipole moment and $\omega$ 
is the frequency. From here, we make further 
approximations: given a distance $2d$ separating the dipoles, we can let $r_1 \approx r +
 d \cos \theta$ and $r_2 \approx  r - d \cos \theta$. Further, since we are assuming $d \ll r$, we 
 approximate $r_1 \approx r_2 \approx r$ in the denominators. The z-component is then
\begin{equation}
E_z = \frac{p_e \,\omega^2}{2\pi}\frac{e^{i\omega r}}{r}\sin^2\theta \cos(\omega d\,\cos\theta)
\end{equation}
Similarly, for the other non-zero field components, we have:
\begin{equation}
E_x = -\frac{p_e \, \omega^2}{4\pi}\frac{e^{i\omega r}}{r}\sin\theta\cos\theta\cos(\omega 
d\, \cos\theta) \,,
\end{equation}
and
\begin{equation}
H_y = -\frac{p_e \, \omega^2}{2\pi}\frac{e^{i\omega r}}{r}\sin\theta\cos(\omega 
d \, \cos\theta)
\end{equation}

The next step is to obtain $P(\theta)$, the power radiated per unit solid angle
in the direction of a unit vector 
$\mathbf{n} = \sin\theta\, \mathbf{\hat{x}} + \cos\theta \,\mathbf{\hat{z}}$. 
From the Poynting vector, we find
\begin{eqnarray}
P(\theta) &=& r^2\, \textbf{n}\cdot(\textbf{E}\times\textbf{H}^{\ast}) 
= r^2\, (\sin\theta\, \mathbf{\hat{x}} + \cos\theta\, \mathbf{\hat{z}})
\cdot  (- E_z H_y^{\ast} \, \mathbf{\hat{x}}  + E_x H_y^{\ast} \, \mathbf{\hat{z}} ) 
	\nonumber \\
 &=& \frac{p_e^2 \,\omega^4}{8\pi^2}\, [ \sin^4\theta\cos^2(\omega d\cos\theta) 
	  + \sin^2\theta\cos^2\theta\cos^2(\omega d\cos\theta)] \nonumber \\
 &=& \frac{p_e^2 \,\omega^4}{8\pi^2}\, [\sin^2\theta\cos^2(\omega d\cos\theta)] \,.
\end{eqnarray}
We next integrate $P(\theta)$ to obtain the total power radiated:
\begin{equation}
P_T = \int_{0}^{2\pi} \int_{0}^{\pi/2} P(\theta)d\Omega \,.
\end{equation}
Let $u = \cos\theta$ and use $p_e = q A$ and $\xi = \omega\, d$ to write
\begin{equation}
P_T  = \frac{p_e^2 \,\omega^2}{4\pi}\int_0^1 (1-u^2)\cos^2(\xi u)du 
 = \frac{p_e^2 \,\omega^2}{96\pi} \,\left\{8+\frac{3}{\xi^3}[-2\xi\cos(2\xi) + \sin(2\xi)]\right\} 
 = \frac{q^2A^2\omega^4}{12\pi}S_T \,,
\end{equation}
where
\begin{equation}
S_T = 1 + \frac{3}{8\xi^3}[-2\xi\cos(2\xi) + \sin(2\xi)] \,.
\end{equation}
This gives us $P_T$, the energy radiated per unit time. 

We can write the energy radiated  per oscillation cycle, $E_c$, as
\begin{equation}
E_c = \frac{2\pi P_t}{\omega} = \frac{1}{6}q^2A^2\omega^3S_T \,.
\end{equation}
The ratio of this quantity to the particle's average kinetic energy is
\begin{align}
\frac{E_c}{\langle E_{\rm kin} \rangle} &= \frac{q^2A^2\omega^3S_T}{ 3m\langle v^2 \rangle} = 
\frac{2 q^2A^2\omega^3S_T}{3 mA^2\omega^2} \\
\frac{E_c}{\langle E_{\rm kin} \rangle} &= \frac{2q^2\omega}{3m}S_T \,.
\end{align}
The function $S_T$ is of order one when $\xi$ is of order one. Then,
inserting the charge and mass values for an electron, as well as our typical
frequency value of $10^{14} Hz$, the estimate comes out to
\begin{equation}
\frac{E_c}{\langle E_{\rm kin} \rangle} \approx 8\times10^{-9}
\end{equation}

Thus, the electron radiates only a few parts per billion of its own kinetic energy per cycle.
The small value of this ratio shows that the electron with our driving field is a weakly damped 
driven oscillator that needs only minimal energy restoration for preservation. The emitted 
radiation is the primary irreducible source of dissipation. This estimate indicates that it is 
reasonable to neglect its dissipative effects on the motion of the paricle.

\subsection{Shot Noise from Photon Emission}
\label{sec:shot}

However, there is another effect arising from the emitted radiation to be considered.
Because the power radiated by the particle consists of discrete photons, there will be a
statistical uncertainty in the momentum lost by the particle. This will lead to an additional
contribution to $\langle \Delta v^2 \rangle$, the velocity variance of the particle. Any 
experiment which seeks to measure the effects of vacuum fluctuations on the variance, 
Eqs.~(\ref{eq:vz2}) and   (\ref{eq:vx2}), will have to contend with this shot noise as a 
background. Let $P_i$ be the average power radiated by the particle in direction~$i$.
Then in time $t$, an energy and magnitude of momentum of $p_i = P_i \, t$ will be radiated
in this direction, corresponding
to a mean number of photons of $N_i =   P_i \, t/\omega$. The statistical uncertainty in this
number is $\sqrt{N_i}$, assuming that the emission of different photons are uncorrelated 
events. This leads to an uncertainty in the $i$-component of the particle's momentum
of order
\begin{equation}
\Delta p_i = \omega\, \sqrt{N_i} = \sqrt{P_i \omega\, t} \,,
\end{equation}
and a variance in the velocity in direction~$i$ of
\begin{equation}
\Delta v_{si}^2 =\frac{ P_i \omega\, t}{m^2}\,,     \label{eq:shot}
\end{equation}
where the ``s"-subscript refers to shot noise.

Now we find the total power radiated in the z-direction. This quantity is
found by projecting onto the $z$-axis, and integrating over a hemisphere:
\begin{eqnarray}
P_z &=& r^2 \int_{0}^{2\pi} \int_{0}^{\pi/2} P(\theta)\cos\theta\, d\Omega 
 = 2\pi r^2 \int_{0}^{\pi/2} P(\theta)\cos\theta\, d(\cos\theta)   \nonumber \\
 &=&  \frac{p_e^2\omega^4}{4\pi}\int_0^1 u(1-u^2)\, \cos^2(\xi u) \, du \,,
\end{eqnarray}
where $u = \cos \theta$, as before. The result is
\begin{equation}
P_z = \frac{1}{64\pi}\frac{p_e^2}{d^4} \;[-3 - 2\xi^2 + (3 - 4\xi^2)\cos(2\xi) 
	+ 2\xi(\xi^3 + 3\sin(2\xi))] \,.
\end{equation}
Next, introduce substitutions for the dipole moment as follows:
\begin{equation}
p_e^2 = q^2A^2 = q^2\left(\frac{qE_0}{m\omega^2}\right)^2 \,.
\end{equation}
We now have
\begin{equation}
P_z = \frac{q^4E_0^2}{64\pi m^2}\frac{S_z}{\xi}
\end{equation}
where,
\begin{equation}
\frac{S_z}{\xi} = \frac{1}{\xi^4}[-3 - 2\xi^2 + (3 - 4\xi^2)\cos(2\xi) + 2\xi(\xi^3 + 3\sin(2\xi))]\,.
\end{equation}
Consequently the mean square velocity in the $z$-direction from shot noise is
\begin{equation}
\Delta v_{sz}^2 =  \frac{q^4E_0^2S_z}{64\pi m^4d}\,t  \,.
\end{equation}
Now compare this effect to that of the electric field fluctuations, using Eq.~(\ref{eq:vz2})
 to write
\begin{equation}
\frac{\langle \Delta v^2_z \rangle}{\Delta v_{sz}^2} = 
\frac{4R_z}{S_z} \,.
\end{equation}
\begin{figure}
\includegraphics[scale=0.5]{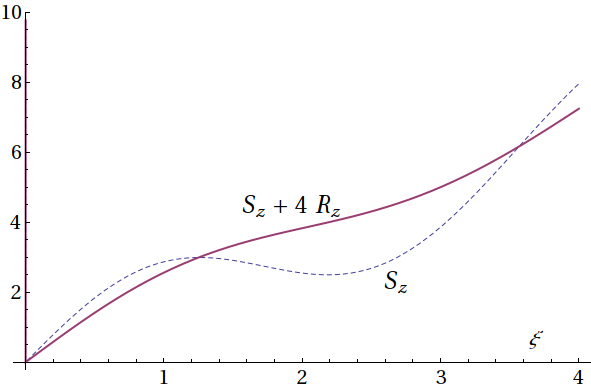}
\caption{ The relative magnitudes of velocity variance in the z-direction with only 
shot noise, $S_z$, and with both shot noise and quantum electric
field fluctuations, $S_z + 4 R_z$.}
\label{fig:shotnoise_z}
\end{figure}
Figure~\ref{fig:shotnoise_z} compares these effects, showing the relative magnitudes of what 
would be seen without and with quantum electric field fluctuations, as a function of $\xi$. 

\begin{figure}
\includegraphics[scale=0.6]{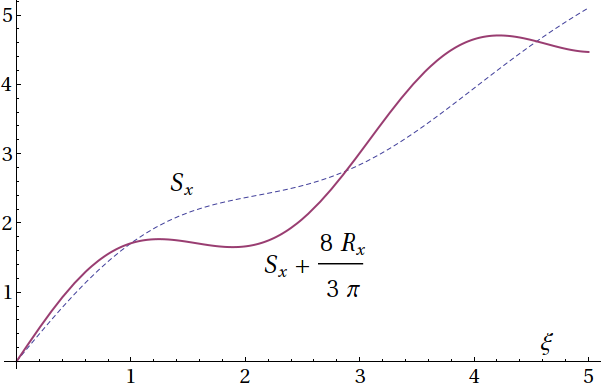}
\caption{ The relative magnitudes of velocity variance in the x-direction with only 
shot noise, $S_x$, and with both shot noise and quantum electric
field fluctuations.}
\label{fig:xgraph}
\end{figure}

We can make a similar calculation for the 
power radiated in the x-direction, and find
\begin{equation}
P_x = r^2 \int_{0}^{2\pi} \int_{0}^{\pi/2} P(\theta)\sin\theta \, d\Omega 
 = \frac{3}{128}\frac{q^4E_0^2}{m^2}\frac{S_x}{\xi} \,.
\end{equation}
Here 
\begin{equation}
S_x =  \xi\,  \left[2\frac{J_2(2\xi)}{\xi^2} + 1 \right] \,,
\end{equation}
and $J_2$ is a Bessel function of the first kind.
We  find the x-direction velocity variance to be
\begin{equation}
\Delta v^2_{sx} = \frac{3}{128}\frac{q^4E_0^2}{m^4d}S_x t\,.
\end{equation}
The ratio of the effect of electric field fluctuations to that of shot noise for the
transverse direction is
\begin{equation}
\frac{\langle \Delta v^2_x \rangle}{\Delta v_{sx}^2} = 
\frac{8R_x}{3 \pi S_x} \,.
\end{equation}
The same graphical comparison as for the z-direction leads to  Fig.~\ref{fig:xgraph}.
We see that the effects of quantum electric field fluctuations and of shot noise are comparable
in order of magnitude when $\xi$ is of order one. However, the sum of the two effects always 
seems to lead to a positive velocity variance. In the limit that $\xi \gg 1$, we find
\begin{equation}
\frac{\langle \Delta v^2_z \rangle}{\Delta v_{sz}^2} \sim 2 \frac{\cos(2 \xi)}{\xi^2} \,,
\end{equation}  
and
\begin{equation}
\frac{\langle \Delta v^2_x \rangle}{\Delta v_{sx}^2} \sim \frac{8\sin(2 \xi)}{3 \pi \xi} \,.
\end{equation}  
 Thus, in the limit of high oscillation frequency or large distance to the mirror, the
 shot noise effect dominates.

 \section{Summary}
  
  In summary, we have presented a model in which charges, such as electrons,
  moving in the quantum electromagnetic vacuum near a mirror may increase
  or decrease their velocity variance. The ultimate energy source is the driving
  field, but the mechanism can be viewed as non-cancellation of anticorrelated
  electric field fluctuations. The effect is a form of squeezing of the particle's
  velocity uncertainty by the electromagnetic vacuum fluctuations.  The most
  striking aspect of this effect is that the mean squared velocity can decrease,
  corresponding to an effective cooling of the charges. Although the effect is
  normally small,  it might be observable.
  
  In our model, we have assumed that the charges move and the mirror remains
  stationary. However, for non-relativistic motion, one would obtain the same result
  if the opposite were true. A rapidly oscillating mirror is more difficult to achieve,
  although rapid electrical switching of the reflectivity of a mirror might be possible,
  and has been explored in the context of the dynamical Casimir effect, the quantum
  emission of photons by a moving mirror~\cite{FD76,FV82}. 
  This effect seems to have been recently
  observed in the context of superconducting circuits~\cite{Wilson11}. Although the
  effect discussed in the present paper involves exchange of kinetic energy between
  charges and a quantum field in the presence of a boundary, rather than quantum creation 
  of photons, it can be viewed as a variant of the dynamical Casimir effect. In the latter
  case, the kinetic energy of the boundary is converted into photons. In the model of
  this paper, it is converted into random motion of a charged particle, but both are
  effects in quantum field theory.
  
   An alternative to switching of a mirror is the use of charges moving near a corrugated
  mirror, as in the Smith-Purcell effect~\cite{SP}. In this configuration, the effect studied
  here should also arise. 
  
  We compared the effects of electromagnetic vacuum fluctuations with shot noise due
  to emission of a finite number of photons. The two effects can be
   of the same order of magnitude,
  but have distinct signatures, so it should be possible to distinguish them experimentally.
   
  The effect studied here is 
  also of interest as an analog model for quantum effects in cosmology. A curved background 
  spacetime can also cause non-cancellation of otherwise anticorrelated fluctuations. Thus
the effect discussed here bears some  relationship
 the effects studied in Refs.~\cite{Bessa09,WKF07,FMNWW10,WHFN11}.

\begin{acknowledgments}
We would like to thank  Jim Babb, Jen-Tsung Hsiang, Akbar Salam, and Roger Tobin 
 for useful discussions. This research was supported in part by the US National Science 
 Foundation under Grant No. PHY-0855360.
\end{acknowledgments}

\end{document}